# THE PERFECT MATCH? A CLOSER LOOK AT THE RELATIONSHIP BETWEEN EU CONSUMER LAW AND DATA PROTECTION LAW

NATALI HELBERGER, FREDERIK ZUIDERVEEN BORGESIUS & AGUSTIN REYNA[*]

**Abstract**
*In modern markets, many companies offer so-called "free" services and monetize consumer data they collect through those services. This paper argues that consumer law and data protection law can usefully complement each other. Data protection law can also inform the interpretation of consumer law. Using consumer rights, consumers should be able to challenge excessive collection of their personal data. Consumer organizations have used consumer law to tackle data protection infringements. The interplay of data protection law and consumer protection law provides exciting opportunities for a more integrated vision on "data consumer law".*

## 1. Introduction

For a long time, consumer law and data protection law belonged to two different worlds. Consumer law is primarily concerned with consumers and their relations with traders of products and services. Consumer law confers mandatory rights on consumers, so as to create a fair legal playing field for economic transactions. Data protection law aims to protect fairness and fundamental rights when personal data are processed. Consumer law deals with fair contracting; data protection law with fair processing.

[*]Natali Helberger and Frederik Zuiderveen-Borgesius are affiliated with the IViR Institute for Information Law of the University of Amsterdam; Agustin Reyna with BEUC, the European Consumer Organisation, Brussels. Views expressed are personal. The authors would like to thank the editors of the CML Rev. and the anonymous reviewers for their constructive suggestions. The authors would also like to thank Damian Clifford and the participants of the Privacy Law Scholars Conference (PLSC2017) for their excellent comments, suggestions and inspiration for further research, in particular: Julie Bril, Tal Zarsky, Lilian Edwards, Christian Svanberg, Judith Rauhofer, Deirdre Mulligan, Julie Cohen, Helen Nissenbaum, Michael Vealle, Jan Whittington, Joseph Tuow, Ron Lee and Joris van Hoboken. The research was made possible in parts through an ERC grant PERSONEWS ERC-2014-STG project, grant no: 638514.

In a digital economy, these worlds start coming together. Many online digital services are no longer offered in exchange for money, but in exchange for personal data. "Paying with data" has become a popular, though misleading, phrase. Data are a critical ingredient of many new "smart" data-driven consumer products and services, and the relevance of collecting and processing data as part of offering services to consumers will only further increase with the proliferation of the Internet of Things.

Despite their different constitutional basis,[1] consumer law and data protection law have moved closer together at the level of EU law and policy-making. Already in 2014, the European Data Protection Supervisor (EDPS) initiated a debate on the interplay between data protection, competition law, and consumer protection in the Digital Economy.[2] With the 2015 proposal for a Directive on certain aspects concerning contracts for the supply of digital content,[3] the issue of services that are rendered in exchange of data has officially entered the EU policy agenda. At around the same time, the General Data Protection Regulation was adopted, replacing the Data Protection Directive and addressing issues that partly concern consumer protection, such as data portability.[4] In data-driven consumer markets, the distinction between consumer law and data protection law is far from clear-cut. With the integration of more and more data into consumer products, many data protection issues also become consumer issues, and vice versa.

There are several ways of dealing with the legal aspects of the larger role of personal data in consumer markets. One approach could, for instance, be to argue in favour of a strict division of tasks between consumer law and data protection law, a division in which all data-related matters fall primarily under data protection law as the area of law that lays down the ground rules for fair data processing.[5] A possible drawback of such an exclusive focus on data protection is that data protection law covers more and more aspects of data-driven consumer markets. Also, there is growing criticism that data protection law alone, with its strong focus on informed consent as a legal basis for data processing in consumer transactions, may not always provide optimal protection of the interests of digital consumers.[6] Another approach is to accept that the relevant legal framework in data-driven consumer markets is not a matter of "either data protection law or consumer law", but that, instead, data protection law and consumer law could apply in parallel, and could ideally complement each other and offer a sufficiently diverse toolbox of rights and remedies to provide a high level of protection of consumers in digital markets. The interplay between consumer law and data protection law, however, is not yet well understood.[7]

---

[1] Data protection is a right in itself under Art. 7 of the EU Charter of Fundamental Rights, while the promotion of consumers' interest is an objective set in Art. 169 TFEU but generally used in the context of the Internal Market competence under Art. 114 TFEU.

[2] EDPS (2014), "Privacy and competitiveness in the age of big data: The interplay between data protection, competition law and consumer protection in the Digital Economy",
<secure.edps.europa.eu/EDPSWEB/webdav/shared/Documents/Consultation/Opinions/2014/14-03-26_competitition_law_big_data_EN.pdf> (all websites accessed 20 June 2017).

[3] In late 2015, the Commission presented the Proposal for a Directive on certain Aspects concerning Contracts for the Supply of Digital Content (COM(2015)634 final). As well as rules about fair contracting in the digital environment, this draft Digital Content Directive introduces consumer rights in relation to the data supplied in exchange of digital products.

[4] General Data Protection Regulation (EU) 2016/679 of the European Parliament and of the Council of 27 Apr. 2016 on the protection of natural persons with regard to the processing of personal data and on the free movement of such data, and repealing Directive 95/46/EC O.J. 2016, L 119/1; Directive 95/46 EC of the European Parliament and of the Council of 24 Oct. 1995 on the protection of individuals with regard to the processing of personal data and on the free movement of such data (Data Protection Directive), O.J. 1995, L 281/31.

[5] This "division of tasks" is apparent in consumer law, as demonstrated e.g. by the suggestion in the draft Digital Content Directive (at 4) that the protection of individuals with regard to the processing of personal data is governed by data protection law, and that the draft directive "should be made in full compliance with that legal framework". See also (on Personalized Pricing): European Commission, Staff Working Document on Guidance on the Implementation/Application of Directive 2005/29/EC on Unfair Commercial Practices", SWD(2016)163 final, at 148.

[6] Zuiderveen Borgesius, *Improving Privacy Protection in the area of Behavioural Targeting* (Kluwer Law International, 2015); Helberger et al., "Online tracking: Questioning the power of informed consent", 14 *Info* (2012), 57-73; McDonald and Cranor, "The cost of reading privacy policies", 4 *Journal of Law and Policy* (2008),

This paper aims to explore the relationship between consumer law and data protection, to identify where the two fields can complement each other, and to highlight some of the challenges. This paper ultimately hopes to lay the basis for moving towards a more integrated vision of "data consumer law". The paper is structured as follows. Section 2 explains how data affect consumers and consumer transactions. Section 3 introduces data protection law. In section 4, we discuss more concretely some areas in which data protection and consumer law could complement each other, with a special focus on informing consumers, so-called "free" services, unfair terms, unfair practices, and consumer vulnerability. In section 5 we mention some caveats. Section 6 concludes.

2.     **Data and consumers**

Personal data have become increasingly important for consumer protection policy. Personal data are economic assets, and are used to develop modern services, to categorize consumers, and to influence consumers.

*Data as an economic asset:* The notion of data as a currency has become commonplace. Many online transactions involve sharing consumer data with sellers, advertisers, and third parties. For instance, companies may ask data from consumers during registration procedures, or may generate data while consumers use the service. Many so-called "free" online services enable companies to collect consumer data.[8] Often, companies use free services to collect data that have little to do with the actual product, but which companies can monetize (e.g. a babyphone app that requests access to pictures and contact lists).

*Data as part of the service:* More and more products and services rely on collecting and processing personal data: health and fitness devices, personal efficiency apps, personal assistants, search engines, social networks, smart TVs, and connected devices in the Internet of Things. Personal data processing is often necessary to offer or improve services. But the fair processing of personal data is also increasingly part of the reasonable expectations that consumers have regarding services and products.

*Data as a means to determine the conditions of the service:* The collection and use of personal data happens during interactions between sellers and consumers. Many companies develop profiles of individual consumers, based on large sets of data. These profiles can be used, for instance, to target advertising, to personalize recommendations, and to customize

---

543-568; Koops, "The trouble with European data protection law", 4 *International Data Privacy Law* (2014), 50-261.

[7]A point that is further underlined by the controversial debate about the drat Digital Content Directive. For example, the tech industry criticized the Commission's draft Digital Content Directive arguing that it threatens innovation in the app developers sector, <euobserver.com/opinion/136202>.

[8]Hoofnagle and Whittington, "Free: Accounting for the costs of the Internet's most popular price", (2014) *UCLA Law Review*, 606-670; Strandburg, "Free fall: The online market's consumer preference disconnect", (2013) *University of Chicago Legal Forum*, 95-172.

products or services. But such profiles can also be used to personalize prices (price discrimination),[9] or to decide whether a consumer can borrow money.[10]

*Data as a means to influence consumer decision-making:* knowledge is power; and so is knowledge about consumers. A company could use its knowledge about consumers to identify and exploit their personal biases and weaknesses to make them take certain decisions.[11] Companies could, for example, calculate the best time to advertise beauty products (in the morning),[12] or which consumer is more likely to cancel a subscription and consider subscribing to a new service.[13]

In sum, personal data have become an integral part of many products, services and consumer transactions, particularly in the online environment, and data shape the relationships between companies and consumers. Therefore, data have become an important concern – not only for data protection but also for consumer policy.

### 3.     Data and data protection law

The main legal instrument to protect privacy and fairness when personal data are processed is data protection law. In the following, we sketch a brief overview of data protection law, and in particular the recently adopted General Data Protection Regulation (GDPR).[14] The GDPR retains the main principles of the 1995 Data Protection Directive, but adds more details and aims to improve compliance and enforcement.[15]

The main objective of data protection law is to realize the fundamental right to data protection, also contained in the EU Charter,[16] and to ensure that the processing of personal data happens lawfully, fairly, and transparently.[17] Data protection law applies as soon as personal data are processed (with certain exceptions).[18] Personal data are defined as "any information relating to an identified or identifiable natural person."[19]

Data protection law grants rights to data subjects - people whose data are processed - and imposes obligations on data controllers and parties that process personal data. A data controller is "the natural or legal person, public authority, agency or other body which, alone or jointly with others, determines the purposes and means of the processing of personal data."[20] Data protection law applies to both the private and the public sector (with some exceptions).[21]

---

[9]White House Report, "Big Data and differential pricing" (White House, 2015), <www.whitehouse.gov/sites/default/files/whitehouse_files/docs/Big_Data_Report_Nonembargo _v2.pdf>; Zuiderveen Borgesius, "Online price discrimination and data protection law", Amsterdam Law School Research Paper No. 2015-32, 1-21, <ssrn.com/abstract=2652665>.

[10]Citron and Pasquale, "The scored society: Due process for automated predictions", (2014) *Washington Law Review*, 1-33.

[11]Calo, "Digital market manipulation", (2014) *The George Washington Law Review,* 995-1051, at 1003.

[12]Rosen, "Is this the grossest advertising strategy of all time?", (2013) *The Atlantic* <www.theatlantic.com/technology/archive/2013/10/is-this-the-grossest-advertising-strategy-of-all-time/280242/>.

[13]Siegel, *Predictive Analytics: The Power to Predict Who Will Click, Lie, or Die* (John Wiley & Sons, 2013), pp. 6-7.

[14] Cited *supra* note 4.

[15]See generally on the GDPR, De Hert and Papakonstantinou, "The new General Data Protection Regulation: Still a sound system for the protection of individuals?", 32 *Computer Law & Security Review* (2016), 179-194.

[16]Art. 1(2) GDPR; Art. 8 Charter of Fundamental Rights of the European Union.

[17]Art. 5(1)(a) GDPR.

[18]See Art. 1 GDPR. See for the personal data definition Art. 4(1) GDPR; for the processing definition Art. 4(2) GDPR. See for the exceptions e.g. Arts. 2 and 85 GDPR.

[19]Art. 4(1) GDPR.

[20]Art. 4(7) GDPR.

[21]See Art. 2 GDPR for the exceptions.

Hence, a data controller is often, but not always, a company or a trader; a data controller can also be an individual or a government agency.

Data protection law operates on the basis of a number of central principles: lawfulness, fairness and transparency; purpose limitation; data minimization; accuracy; storage limitation; integrity and confidentiality;[22] and accountability.[23] In summary, the principles lead to the following requirements:

(a) Personal data may only be processed "lawfully, fairly and in a transparent manner in relation to the data subject."[24] Fairness requires transparency; for instance, secret data collection is almost never allowed.[25] Data controllers must comply with detailed transparency requirements, and must provide data subjects with all information that is necessary to ensure fair and transparent data processing.[26]

(b) The purpose limitation principle says that personal data may only be "collected for specified, explicit and legitimate purposes and not further processed in a manner that is incompatible with those purposes."[27] Hence, data collected for goal X may generally not be used for goal Y. Scandals about privacy violations or unfair data processing often happen when personal data are used for goals other than people expected.

In data protection law, a purpose can only be "legitimate" if the data controller can base the processing on a legal basis. The GDPR lists six possible legal bases. For the private sector, the most relevant legal bases are "necessity for contract performance", the "legitimate interests" provision, and the data subject's consent.[28] Data controllers thus do not always need to obtain the data subject's consent to be able to process personal data. In some cases, companies can process personal data on the legal basis "necessity for contract performance". A data controller may process personal data if the processing is necessary for the performance of a contract with the data subject.[29] Say a company offers a smart phone app that shows the local weather report. Let us assume that a consumer who buys the app enters into a contract with the company. If the user does not want to enter manually the city where he or she is, the app needs access to the phone's location data to show the local weather report. Hence, the processing of some personal data (location data) is "necessary" to offer the localized weather report function. Therefore, the company is allowed to process those location data for the weather report, because the processing is "necessary" to perform the contract.

Sometimes a company is allowed to process personal data based on the legitimate interest provision, also called the balancing provision. Data controllers can process personal data on the basis of that provision if the processing is necessary for the legitimate interests pursued by the controller, except where such interests are overridden by the interests or fundamental rights of the data subject.[30] For instance, an app provider may process some user data (analytics data) to improve the app. Such data processing for analytics can, under certain circumstances, be based on the legitimate interest provision.

If a data controller cannot base the processing on necessity for contract performance or on the legitimate interest provision (or on another legal basis[31]), the controller must ask the data subject for consent to be able to process personal data.[32] If the app provider wants to share the user's data with third parties, for instance for targeted marketing, consent is typically the only appropriate legal basis.[33] In sum, data controllers may only process personal data if they have a legal basis, such as consent, for processing.

---

[22]Art. 5(1)(a)-(f) GDPR.
[23]Art. 5(2) GDPR.
[24]Art. 5(1)(a) GDPR.
[25]There are some exceptions; see Art. 14(5) GDPR.
[26]Arts. 13(2) and 14(2) GDPR.
[27]Art. 5(1)(b) GDPR.
[28]Art. 6(1)(a), 6(1)(b), and 6(1)(f) GDPR.
[29]Art. 6(1)(b) GDPR.
[30]Art. 6(1)(f) GDPR.
[31]In some circumstances, a data controller could rely on necessity for a legal requirement (Art. 6(1)(c) GDPR) or on necessity to protect the vital interests of the data subject (Art. 6(1)(d) GDPR).

(c) The data minimization principle is that data should be "adequate, relevant and limited to what is necessary in relation to the purposes for which they are processed."[34] Hence, collecting or storing disproportionate amounts of personal data is not allowed, not even after the data subject's consent.[35]

(d) The accuracy principle says that personal data must be accurate and, where necessary, kept up to date. Data controllers must take every reasonable step to ensure that personal data that are inaccurate, having regard to the purposes for which they are processed, are erased or rectified.[36]

(e) The storage limitation principle prohibits storing data for unreasonably long periods. Personal data must be "kept in a form which permits identification of data subjects for no longer than is necessary for the purposes for which the personal data are processed."[37] Under certain conditions, personal data may be stored longer, for example for statistical purposes.[38]

(f) The integrity and confidentiality principle concerns data security. Personal data must be processed in a manner that ensures appropriate security of the personal data.[39]

(g) The accountability principle emphasizes that the data controller is responsible for, and must be able to demonstrate compliance with, data protection law.[40]

Data protection law aims to protect fairness and fundamental rights when personal data are processed – it does not merely aim to protect privacy.[41] An important goal is protecting people against abuse of information asymmetry.[42] Many data protection provisions aim to increase the transparency of processing activities. Through such provisions, data protection law aims to improve the position of the data subject in relation to data controllers. Data protection law grants data subjects several rights. Data subjects have the right to access their data;[43] they have the right to obtain information regarding processing purposes, the categories of data concerned, and the recipients to whom the data are disclosed.[44] Data subjects have the right to rectify, erase or block data, if the processing does not comply with the GDPR's provisions, for example when data are incomplete or inaccurate.[45] Data subjects also have a right to data portability: to move their own personal data from one service provider to another.[46]

---

[32] Art. 6(1)(f) GDPR. The GDPR has strict rules on when consent is "freely given" and thus valid. See Art. 7 GDPR.

[33] "Consent should be required, for example, for tracking and profiling for purposes of direct marketing, behavioural advertisement, data-brokering, location-based advertising or tracking-based digital market research", Art. 29 Working Party 2014, Opinion 06/2014 on the notion of legitimate interests of the data controller under Art. 7 of Directive 95/46/EC (WP 217), 9 Apr. 2014, at 47.

[34] Recital 39 GDPR; Art. 5(1)(c) GDPR.

[35] National courts have ruled that data processing can be unlawful because it is disproportionate, even though the data subject has consented to the processing. Hoge Raad [Dutch Supreme Court], 9 Sept. 2011, NL:HR:2011:BQ8097 (Santander). See for an English summary, Valgaeren and Gijrath, "Supreme Court interprets Dutch Privacy Act in accordance with Article 8 ECHR" (22 Nov. 2011) <www.lexology.com/library/detail.aspx?g=01e9a9c2-3876-4d69-8fb5-85fab7b612ab>. See, in Poland, Naczelny Sąd Administracyjny [Supreme Administrative Court], 1 Dec. 2009, I OSK 249/09 (Inspector General for Personal Data Protection), English translation: <www.giodo.gov.pl/417/id_art/649/j/en/>. See also Rouvroy and Poullet: "even in case of unambiguous consent … it may be possible to declare the processing illegitimate if that processing is disproportionate." (Rouvroy and Poullet, "The right to informational self-determination and the value of self-development: Reassessing the importance of privacy for democracy" in Gutwirth et al. (Eds.), *Reinventing Data Protection?* (Springer, 2009), p. 73). See also Bygrave and Schartum, "Consent, proportionality and collective power" in Gutwirth et al. (Eds.); Gellert and Gutwirth, "The legal construction of privacy and data protection", 29 *Computer Law & Security Review* (2013), 522, 527.

[36] Art. 5(1)(d) GDPR.

[37] Art. 5(1)(e) GDPR.

[38] Art. 5(1)(e) GDPR.

[39] Art. 5(1)(f) GDPR. See on security Arnbak, *Securing Private Communications* (Kluwer Law International, 2016).

[40] Art. 5(1)(2) GDPR.

[41] Privacy and private life are each mentioned only once in the GDPR. Private life is mentioned in Recital 4; privacy is mentioned in the footnote to Recital 173.

[42] See De Hert and Gutwirth, "Privacy, data protection and law enforcement: Opacity of the individual and transparency of power" in Claes, Duff and Gutwirth (Eds.), *Privacy and the Criminal Law* (Intersentia, 2006);

If a data controller breaches data protection law, data subjects have several possibilities. They can file a complaint with the national Data Protection Authority,[47] and can also go to a court if a data controller breaches data protection law.[48] They can also mandate a not-for-profit organization to lodge a complaint on their behalf.[49] Data Protection Authorities monitor compliance, examine potential breaches of data protection law, and can impose sanctions if appropriate.[50]

## 4. Data and consumer law

Unlike data protection law, the rationale of consumer law is less clearly linked to the protection of a fundamental right. Consumer law aims to set the basic rules for the bargaining game between "persons acting as consumer in the market place and their counter-parts, the businesses."[51] In addition, under the EU Charter of Fundamental Rights, consumers are entitled to a high level of consumer protection.[52] Insofar, two distinct rationales underlie European consumer law: to empower consumers as sovereign market actor, giving them the rights and information necessary to act in that role, and to protect consumers in situations where they are the weaker party in commercial dealings, and not able to take the protection of their rights, (economic) interests and safety into their own hands.[53] Consumer law has been traditionally associated with the protection of consumers in the "physical" world. The first Community consumer programme, in 1975, sought to co-ordinate the various national policies to protect consumers interest when purchasing goods.[54] However, the EU aims to make consumer law digital-proof. Starting with the 2011 Consumer Rights Directive,[55] policymakers in the field of consumer law pay increasing attention to the concerns of the digital consumer.

---

Zuiderveen Borgesius, *Improving Privacy Protection in the area of Behavioural Targeting* (Kluwer Law International, 2015), Ch. 4, section 4, and Ch. 7.
[43] Art. 15 GDPR.
[44] Art. 13 GDPR.
[45] Arts. 16 and 17 GDPR.
[46] Art. 20 GDPR.
[47] Art. 78 GDPR.
[48] Art. 79 GDPR.
[49] Art. 80 GDPR.
[50] Art. 58 (1) and (2) GDPR.
[51] Wilhelmson, "Consumer law and the environment: From consumer to citizen", 21 *Journal of Consumer Policy* (1998), 45, at 46.
[52] Art. 38, EU Charter of Fundamental Rights.
[53] European Parliament, Consumer Protection in the EU, PE 565.904, Sept. 2015.
[54] Council Resolution of 14 Apr. 1975 on a preliminary programme of the European Economic Community for a consumer protection and information policy, O.J. 1975, C 92/01.
[55] Directive 2011/83/EU of the European Parliament and the Council of 25 Oct. 2011 on consumer rights, amending Council Directive 93/13/EEC and Directive 1999/44/EC and repealing Council Directive 85/577/EEC and Directive 97/7/EC (Consumer Rights Directive), O.J. 2011, L 304/64. See also the preparatory works to the Directive, Loos, Helberger, Guibault, Mak, Pessers, Cseres, "Analysis of the applicable legal frameworks and suggestions for the contours of a model system of consumer protection in relation to digital content contracts",

Parallel to that, a group of experts drafted provisions that were to form the basis for the European Commission's proposal for a Regulation on a Common European Sales Law.[56] Both the Consumer Rights Directive and the proposed Common European Sales Law address contracts for the provision of digital content.[57]

So far, personal data have played only a small role in the process of amending the consumer law framework to meet the needs of the digital economy. Rather, the EU has focused on adjusting traditional consumer law instruments to digital services. In December 2015, however, the Commission published a proposal for a Directive on certain Aspects concerning Contracts for the Supply of Digital Content (Digital Content Directive).[58] This replaces the Common European Sales Law proposal and aims to harmonize certain rules for the supply of digital content, and set a high level of consumer protection throughout the EU. The directive deals explicitly with data – a novelty in EU consumer law.

The draft Digital Content Directive acknowledges the economic reality that many digital services are offered not in exchange for a monetary payment but in exchange for personal data. It appears that the directive aims to adapt legal reality to that economic reality.[59] For its relationship with data protection law, the draft directive notes that the protection of individuals with regard to the processing of personal data is governed by data protection law, and that the draft directive "should be made in full compliance with that legal framework".[60] However, as discussed below, the relationship between consumer law and data protection law is more complicated.

The following sections look more closely at the question what consumer law could add to the existing level of protection under data protection law; they address informing consumers about so-called "free" services, unfair terms, unfair practices, and consumer vulnerability.

### 4.1. *Informing consumers*

One feature that unites consumer law and data protection law is the pivotal role of information as a means to mitigate information asymmetries and to empower the individual.[61] Data protection law, in particular the new GDPR, devotes considerable attention to the question of how to inform consumers in such a way that they are able to take informed decisions about how companies deal with their data. The principle of informed consent is a key legal basis for the lawful processing of personal data.[62] In addition, the GDPR stipulates an impressive list of

---

Final report to the European Commission (University of Amsterdam, 2011). Helberger, Loos, Guibault, et al., "Digital content contracts for consumers", 36 *Journal of Consumer Policy* (2013), 37-57.

[56]Proposal for a regulation of the European Parliament and of the Council on a Common European Sales Law (CESL). COM(2011)635 final.

[57]See Reyna, "What place for consumer protection in the Single Market for digital content? Reflections on the European Commission's optional regulation policy", 2 *Revue Européenne de Droit de la Consommation* (2014), 333-362.

[58]COM(2015)634 final. Cited *supra* note 3.

[59]Arts. 3 and 13 of the proposed Digital Content Directive.

[60]Proposed Digital Content Directive, at 4.

[61]Fritsch, Wein and Ewers, *Marktversagen und Wirtschaftspolitik*, 7th edn., (Vahlen, 2007), p. 305. Grundmann and Kerber, "Information intermediaries and party autonomy: The example of security and insurance markets", in Grundmann, Kerber and Weatherill (Eds.), *Party Autonomy and the Role of Information in the Internal Market,* (De Gruyter, 2001), pp. 264-310, p. 269. But also Fung, Graham, and Weil, "The political economy of transparency: What makes disclosure policies sustainable?", (2014) KSG Working Paper No. RWP03-039, <papers.ssrn.com/sol3/papers.cfm?abstract_id=384922Helberger> and Helberger, "Form matters: Informing consumers effectively", final report for BEUC (University of Amsterdam, 2013) <www.ivir.nl/publicaties/download/Form_matters.pdf>. Zuiderveen Borgesius, *Improving Privacy Protection in the Area of Behavioural Targeting* (Kluwer Law International, 2015).

[62]Art. 6(1)(a) GDPR.

items that individuals should be informed about, including the processing purpose, the contact details of the controller, and all other information that is necessary to ensure the fairness of the personal data processing.[63] The GDPR also makes stipulations regarding the form in which the information should be given: concise, easily accessible, and easy to understand, using clear and plain language.[64]

The GDPR contains more detailed transparency obligations than the EU consumer protection directives. The GDPR's preamble, for example, calls for visualization where appropriate,[65] and for "standardized icons in order to give in an easily visible, intelligible and clearly legible way a meaningful overview of the intended processing."[66] Elsewhere the GDPR mentions certifications, seals, and marks as a means of enhancing transparency.[67] Special attention is required for the form in which the information should be given to children, namely in "such a clear and plain language that the child can easily understand."[68] The GDPR's preamble recommends providing information in electronic form (e.g. through a website), especially in situations "where the proliferation of actors and the technological complexity of practice make it difficult for the data subject to know and understand whether, by whom and for what purpose personal data relating to him or her are being collected, such as in the case of online advertising."[69] The drafters of the GDPR appear to have realized that merely informing people is not enough to empower them, particularly if the information is provided in a format that is neither attractive nor clear.

The 2011 Consumer Rights Directive adds few personal data-related information requirements. However, the Directive requires companies to inform the consumer about the functional aspects of the digital content. The functionality refers to the ways in which the digital content can be used, including "tracking of consumer behaviour".[70] According to the European Commission, the functionality also refers to whether personalization happens.[71]

The true added benefit of consumer law to inform consumers about personal data use could be in the extra level of flexibility and attention to the individual context that consumer law affords. While data protection law prescribes in detail which information about data processing consumers should be given, and in which form, consumer law, and in particular the Unfair Commercial Practices Directive, Directive 2005/29, provides extra flexibility. Arguably, the material information that traders are required to disclose according to the Unfair Commercial Practices Directive must include information about collecting and monetizing personal data. According to Article 6 of the Unfair Commercial Practices Directive, "a commercial practice shall be regarded as misleading if it contains false information and is therefore untruthful or in any way, … and in either case causes or is likely to cause him to take a transactional decision that he would not have taken otherwise"; the notion of "transactional decision" by the consumer must be interpreted broadly, to also cover the decision to enter into a contract.[72] Before being able to make an informed transactional decision in the sense of the Unfair Commercial Practices Directive, consumers must be aware of whether a company will collect their personal data and for what purposes.

---

[63]See Recitals, 20, 48 and Art. 14 GDPR.
[64]Similar requirements can be found in traditional consumer contract law. For instance, the Unfair Contract Terms Directive requires standard contract terms to be "drafted in plain, intelligible language" (Directive 93/13 on unfair terms in consumer contracts, O.J. 1993, L 95/29). The GDPR refers to the Unfair Contract Terms Directive in Recital 42.
[65]Recital 48 GDPR.
[66]Art. 12(7)(8) and Recitals 60 and 166 GDPR.
[67]Recital 77 GDPR.
[68]Recital 40 GDPR. For consumer law, such special requirements regarding the form in which information is communicated to children are most likely to flow from Art. 3(3) of the Unfair Commercial Practices Directive, and here more specifically the provisions about vulnerable consumers.
[69]Recital 58 GDPR.
[70]Arts. 5 (1)(g), 6 (1)(r) and Recital 19 of the Consumer Rights Directive, Directive 2011/83.
[71]Commission, DG Justice Guidance Document concerning Directive 2011/83/EU on consumer rights (2014), at 76: "As appropriate to the product, the following information should be given: … Any conditions for using the product not directly linked to interoperability, such as: a. tracking and/or personalization" <ec.europa.eu/justice/consumer-marketing/files/crd_guidance_en.pdf>.

Information that consumers need to be informed about if they are not to be misled may include: whether the prices they see are personalized; whether more personal data are captured than necessary to provide the service; and whether those data are monetized or shared with other companies.[73] Much will depend on the actual situation: whether having that information would cause the consumer to take a transactional decision they would not have taken otherwise. If this were the case, not providing that information could be considered an unfair commercial practice. For instance, suppose a consumer bought a smart watch and, for the consumer, it was critical that the smart watch provider does not share the consumer's personal data with third parties. The provider failed to inform the consumer that it shares the consumer's data with third parties, and the provider does in fact share those data in a way that is against the reasonable expectations of the consumer. Such unexpected data sharing could not only constitute a violation of data protection law, but also constitute an unfair commercial practice. Similarly, if consumers are largely opposed to online price discrimination, a trader must reasonably assume that consumers want to be informed about the fact that the price is different for different categories of consumers.

Another benefit of extending the scope of consumer law to data-related issues lies in giving consumers concrete rights against sellers if information obligations are violated. If a data controller breaches data protection law's information obligations, the processing may become unlawful. That unlawfulness, however, says little about the consequences for a possible contractual relationship between seller and consumer. For example: what is the fate of the contract for the purchase of a smart TV where the supplier has failed to inform the consumer about the fact that the device can collect information on individual viewing behaviour and preferences? If consumers are not happy with the data collection and refuse to give consent, can they also return the TV set and demand a less privacy-invasive model? Consumer protection law could fill that gap.

For example, according to the Consumer Rights Directive, the pre-contractual information automatically becomes part of the contract, with the consequence that contracts can be scrutinized under consumer law.[74] Hence, a deviation from what has been promised under the contract (for example that personal data are not shared with third parties) could be seen as a breach of contract. This would entitle the application of national contract law remedies such as contract termination or damages. At EU level, the harmonization of remedies for breaches of information requirements is limited. For example, the Consumer Rights Directive penalizes traders who do not disclose the existence of additional costs that come on top of the advertised price, but it remains silent if the trader does not provide information concerning the main characteristic of the product.[75] If a trader fails to disclose such information, the remedy would have to be found within the applicable national consumer law regime.[76]

---

[72] Directive 2005/29/EC of the European Parliament and of the Council of 11 May 2005 concerning unfair business-to-consumer commercial practices in the internal market, O.J. 2005, L 149. See Commission Guidance on Directive 2005/29/EC, cited *supra* note 5, at 37, with reference to Case C-281/12, *Trento Sviluppo srl, Centrale Adriatica Soc. Coop. Arl* v. *Autoritate Garante della Concorrenza e del Mercato*, EU:C:2013:859, paras. 36 and 38: "any decision directly related to that decision".
[73] See in this sense Commission guidance, previous footnote, at 27 and 146 et seq.
[74] Art. 6(5) of the Consumer Rights Directive.
[75] Art. 6(6).
[76] See Twigg-Flesner, "Information duties" in Schulte-Nölke (Ed.), "EC consumer law compendium: Comparative analysis" (2007), at 734 <http://ec.europa.eu/consumers/archive/cons_int/safe_shop/acquis/comp_analysis_en.pdf>

A breach of an information requirement could also be seen as an unfair commercial practice.[77] Depending on national law, failing to provide information or providing misleading information could lead to fines or making the contract void. Hence, consumer law could provide consumers with concrete contract law remedies in case of breach of transparency requirements.

Finally, under the draft Digital Content Directive, information provided plays a prominent role in determining whether the digital content product is in conformity with the contract.[78] For example, an app that collects and shares more information with third parties than is specified by the provider, or fails to implement the necessary security safeguards, might not conform to the law; as a result, consumers would have the right to a repair or a refund.[79] Data protection law could also influence consumers' reasonable expectations under the draft Digital Content Directive.[80] For example, based on data protection law, consumers can reasonably expect that companies do not use their personal data for new purposes at random, as such new uses would breach the data protection law principle of purpose limitation.[81] Ideally, the provisions in consumer law could act as an additional incentive for companies to inform consumers clearly about their personal data practices.

It is not suggested here that informing consumers is a panacea. Information requirements have limited potential to empower consumers. Scholars from various disciplines agree that information requirements are not a solution for everything.[82] It is only a slight exaggeration to say: consumers don't read information; if they read, they don't understand; if they understand, they don't act.[83] Some scholars speak of the "failure of mandated disclosure".[84] Nevertheless, well-designed and user-friendly information could help consumers.[85] And the media can report on a company's terms and conditions; sometimes companies react to such media reports.[86] In addition, information could help regulators to examine company practices.[87] In sum, information requirements do not solve all problems, but they can be useful and lead to more market transparency.

---

[77] The situation was similar under the Distance Selling Directive (Directive 97/7) (the predecessor of the Consumer Rights Directive).

[78] Art. (1)(a) of the proposed Digital Content Directive states that in order to conform with the contract, the digital content shall, where relevant: "be of the quantity, quality duration and version and shall possess functionality, interoperability and other performance features such as accessibility, continuity and security, as required by the contract, including in any pre-contractual information which forms integral part of the contract".

[79] Arts. 6 and 12 of the draft Digital Content Directive.

[80] See Art. 6 (1)(a) of the draft Digital Content Directive.

[81] Art. 5(1)(b) GDPR.

[82] See e.g. in the context of consumer law: Howells, "The potential and limits of consumer empowerment by information", 32 *Journal of Law and Society* (2005), at 349. In the context of data protection law, Acquisti and Grossklags, "What can behavioral economics teach us about privacy?" in Acquisti et al. (Eds.), *Digital Privacy: Theory, Technologies and Practices* (Auerbach Publications, 2007), pp. 363-378; Zuiderveen Borgesius, "Behavioural sciences and the regulation of privacy on the Internet" in Sibony and Alemanno (Eds.), *Nudging and the Law: What can EU Law Learn from Behavioural Sciences?* (Hart Publishing, 2015), pp. 179-207.

[83] Ben-Shahar and Schneider arrive at a similar conclusion on the regulatory technique of mandated disclosure of information in general, Ben-Shahar and Schneider, "The Failure of mandated disclosure", 159 *University of Pennsylvania Law Review* (2011), at 665.

[84] Ibid.

[85] See Calo, "Against notice skepticism in privacy (and elsewhere)", 87 *Notre Dame Law Review* (2012), 1027-2261; Helberger, op. cit. *supra* note 61.

[86] E.g., after attention in the press, Facebook offered people a way to opt-out of their "Beacon" service. See Debatin et al., "Facebook and online privacy: Attitudes, behaviors, and unintended consequences", 15 *Journal of Computer-Mediated Communication* (2009), at 83.

[87] Van Alsenoy, Kosta and Dumortier, "Privacy notices versus informational self-determination: Minding the gap", 28 *International Review of Law, Computers & Technology* (2014), 185-203; Hintze, "In defense of the long privacy statement", *Maryland Law Review*, forthcoming, <ssrn.com/abstract=2910583>.

In conclusion, both consumer law and data protection law partly rely on empowering individuals to make informed choices by providing information requirements for companies; they can usefully complement each other. Consumer law adds extra flexibility to the prescribed list of information requirements in data protection law, and also provides additional remedies in case of breach of information obligations.

4.2. *So-called "free" services*

Consumer law has not been applied much, so far, to so-called "free" services, such as apps and websites for which consumers do not pay with money. So far, services that are not rendered against a monetary price will often fall outside the scope of consumer law.[88] As a result, consumers who receive services in exchange for data or attention are entitled to a lower level of protection than consumers that pay money for the service, even if the service is the same. This differentiation seems unfair, taking into account the economic value that personal data have in digital markets.

But the Unfair Commercial Practices Directive and the draft Digital Content Directive contain provisions that could be applied to such "free" services. According to the Annex of the Unfair Commercial Practices Directive, it is considered unfair to "[describe] a product as 'gratis', 'free', 'without charge' or similar if the consumer has to pay anything other than the unavoidable cost of responding to the commercial practice and collecting or paying for delivery of the item."[89] The provision is broad enough to cover the payment of non-monetary forms of remuneration, seeing the lack of a direct reference to notions such as "money" or "price" in the provision. This interpretation seems to be confirmed by the European Commission: "The marketing of such products as 'free' without telling consumers how their preferences, personal data and user-generated content are going to be used could in some circumstances be considered a misleading practice."[90] Moreover, according to the Unfair Commercial Practices Directive, it is a misleading omission if a company fails to identify the commercial intent of a commercial practice (if not already apparent from the context).[91] Hence, a company could breach that provision if it fails to inform consumers that its "free" app captures the consumer's personal data.

The draft Digital Content Directive goes a step further and stipulates concrete rights of consumers of so-called "free services". A controversial innovation in the draft Digital Content Directive is the inclusion of digital services that are provided in exchange for data.[92] Such services are often and misleadingly referred to as "free", while the provider captures personal data from users as a form of payment. Many people may not realize that such "free" services rely on gathering personal data. Even if people realize their personal data are captured, it is questionable whether they can assess whether the amount of data that a company collects (as a form of counter-performance) is fair. People rarely know which data about them are captured, how those data will be used[93] and what the value is of those data.[94]

---

[88] E.g., according to Art. 1 and 2 (5)(6) of the Consumer Rights Directive, contracts for the supply of goods or services covered by the Directive are those " under which the trader supplies or undertakes to supply a service to the consumer and the consumer pays or undertakes to pay the price thereof".
[89] No. 20 of the Annex to the Unfair Commercial Practices Directive
[90] Commission guidance cited *supra* note 72, at 97.
[91] Art. 7(2) Unfair Commercial Practices Directive.
[92] For a discussion see e.g. Metzger, "Data as counter-performance: What rights and duties do parties have?", 8 *Journal of Intellectual Property, Information Technology and E-Commerce Law* (2017), 2-8; Langhanke and Schmidt-Kessel, "Consumer data as consideration", 4 *Journal of European Consumer and Market Law* (2015), 218-223; Loos and Mak, "Remedies for buyers in case of contracts for the supply of digital content", Report for the European Parliament, 2012; Jacquemin, "Digital content and sales or services contracts under EU law and Belgian/French law", 8 *Journal of Intellectual Property, Information Technology and E-Commerce Law* (2017), 27-38; Wendehorst, "Sale of goods and supply of digital content – two worlds apart?", Report for the European Parliament (Juri Committee), Feb. 2016.

Informing consumers about the nature and value of the counter-performance (usually a price) to enable them to make informed purchasing decisions is a cornerstone of consumer law. For example, according to the Consumer Rights Directive, consumers must be informed in advance about "the total price of the goods or services inclusive of taxes."[95] However, this requirement is limited to monetary payments.[96] At the time of debating the Commission's proposal for the Consumer Rights Directive, the European Parliament and the Council did not want to apply this Directive to contracts covering digital content not supplied in exchange for a monetary payment.[97] The German Government indicated that the Consumer Rights Directive should apply only if consumers pay a fee.[98]

The Commission directorate responsible for justice and consumers (DG JUSTICE) gave a broad interpretation of the scope of the Consumer Rights Directive. DG JUSTICE indicated that "contracts for online digital content are subject to the Directive even if they do not involve the payment of a price by the consumer."[99] However, situations in which people access online services without express contractual agreement are excluded.[100] Hence, contracts (for the supply of digital content in exchange of data) that are concluded by tacit agreement would escape the application of the Consumer Rights Directive. Because of that limitation, many "free" online services would be excluded from the scope of the Consumer Rights Directive. It is debatable whether people enter a contract with a website if they merely visit that website. Suppose a website publisher allows dozens of other companies to collect data about its website visitors, for instance through tracking cookies (a common situation on the web).[101] As the website visitors did not expressly enter into a contract with the publisher, the Consumer Rights Directive would not apply. DG JUSTICE concludes: "In itself, access to a website or a download from a website should not be considered a 'contract' for the purposes of the Directive."[102]

---

[93] Acquisti and Grossklags, op. cit. *supra* note 82.
[94] Hoofnagle and Whittington, op. cit. *supra* note 8; Strandburg, op. cit. *supra* note 8.
[95] Art. 5(1)(c) Consumer Rights Directive.
[96] As can be concluded from the wording of Art. 5(1)(c) of the Consumer Rights Directive: "the total price of the goods or services inclusive of taxes, or where the nature of the goods or services is such that *the price cannot reasonably be calculated in advance, the manner in which the price is to be calculated, as well as, where applicable, all additional freight, delivery or postal charges or, where those charges cannot reasonably be calculated in advance, the fact that such additional charges may be payable*" (our emphasis).
[97] E.g., during the Belgian Presidency of the Council, it was suggested that *"... the downloading of digital content by a consumer should be regarded as a service contract which falls within the scope of this directive [CRD] ..."*, while a service contract was defined as a contract in exchange for a price: suggestion for recital (10d) in Meeting Document from the Presidency to the Working Party on Consumer Protection and Information of 29 Oct. 2010, Reference Number: 1747/10. In the final general approach of the Council of 10 Dec. 2010, digital content was included in the directive but as a sui generis category without a clear scope: "(10d) Digital content, such as computer programs, games or music that is not burned on a tangible medium is not considered as tangible items. It is thus not considered as a good within the meaning of this Directive. On the contrary, media containing digital content such as CD/DVD are tangible items and are thus considered as goods within the meaning of this Directive. The downloading of digital content by a consumer from Internet should be regarded, for the purpose of this Directive, as a contract which falls within the scope of this Directive, but without a right of withdrawal. The Commission should examine the need for harmonized detailed provisions in this respect and submit, if necessary, a proposal for addressing this matter.", Reference Number: 16933/10.
[98] Gesetzentwurf der Bundesregierung, "Entwurf eines Gesetzes zur Umsetzung der Verbraucherrechterichtlinie und zur Änderung des Gesetzes zur Regelung der Wohnungsvermittlung", 6 March 2013 <dipbt.bundestag.de/dip21/btd/17/126/1712637.pdf>.
[99] DG Justice Guidance on Directive 2011/83, cited *supra* note 71.
[100] Ibid., at 64.
[101] See: Altaweel, Good & Hoofnagle, "Web Privacy Census. Technology Science", 2015121502. December 15, 2015 <https://techscience.org/a/2015121502>.

Contrary to the Consumer Rights Directive, the draft Digital Content Directive states that the notion of price concerns the exchange of money. The draft Digital Content Directive sees providing data as a separate counter-performance by consumers: "This Directive shall apply to any contract where the supplier supplies digital content to the consumer or undertakes to do so and, in exchange, a price is to be paid *or the consumer actively provides counter-performance other than money in the form of personal data or any other data.*"[103] Hence, the draft Digital Content Directive distinguishes (i) the "price" from (ii) "data" that a consumer supplies as counter-performance. The consideration of data as a counter-performance adds a new dimension in the application of the fairness control mechanism of consumer law. Including data as a counter-performance could open up the application of the provisions about unfair contracts and possibly consumer sales law. For example, suppose a consumer downloads a torch app that collects all kinds of information (contact list, location data, IP address, etc.) that are not related to the functionality of the app itself.[104] Such a transaction could create an unfair balance in the rights and obligations of parties, and thus be unlawful according to the provisions of contract law. The fact that the Digital Content Directive accepts data as a counter-performance may pave the way for applying consumer law to "free" services more generally, for instance in the context of the Consumer Rights Directive, the Unfair Contract Terms Directive,[105] or the Unfair Commercial Practices Directive.

Hence, consumer law has made first steps towards improving the legal standing of consumers of "free" services, even though several aspects of the data-as-counter-performance approach in the draft Digital Content Directive are unclear. For example, the supply of data must be considered as a counter-performance in the draft directive. If a trader explicitly asks a consumer (data subject) to consent to personal data collection for targeted marketing, the data disclosed by the consumer could probably be seen as a counter-performance. The draft Digital Content Directive appears to assume that the consumer gives consent for personal data processing.[106] But the draft Directive is not clear on that point, or on how consent under data protection law and consumer law relate to each other.[107] Moreover, as noted in section 3 above, consent is merely one of the possible legal bases in data protection law.[108] It is unclear how the draft Digital Content Directive interacts with data protection law if a controller relies on another legal basis than the data subject's consent.

One may wonder to what extent it is in the interest of traders to characterize the provision of data as a counter-performance if that means that the contract would fall under the draft Digital Content Directive. It appears that the draft Directive would apply to "free" services only if the trader explicitly acknowledges that it regards user data it collects as a counter-performance. Without such an explicit statement, much will probably depend on circumstantial evidence, such as whether the consumer would be able to use the service also without agreeing to the collection of their personal data.[109]

---

[102]Ibid., p. 64: "In itself, access to a website or a download from a website should not be considered a 'contract' for the purposes of the Directive."
[103]Art. 3(1) of the proposed Digital Content Directive (emphasis added). See also Art. 2(6) of the proposed Digital Content Directive
[104]See Federal Trade Commission, "FTC approves final order settling charges against Flashlight App creator", 9 Apr. 2014, <www.ftc.gov/news-events/press-releases/2014/04/ftc-approves-final-order-settling-charges-against-flashlight-app>.
[105]Council Directive 93/13/EEC on unfair terms in consumer contracts, O.J. 1993, L 95/29.
[106]In this sense also Metzger, op. cit. *supra* note 92, at 5.
[107]See extensively on this point Metzger, ibid., and also section 5 *infra*. See also EDPS, Opinion 4/2017 on the Proposal for a Directive on certain aspects concerning contracts for the supply of digital content, March 2017, paras. 49 et seq., and particularly para 54.
[108]See section 3 *supra*.

The draft Digital Content Directive's wording would also lead to difficult definition problems. The draft Directive applies to a contract where the consumer "actively" supplies data. The word "actively" narrows the scope of the draft Directive. For example, if a consumer visits a website, and the website allows twenty marketing companies to collect data about the consumer with tracking cookies, the consumer does not seem to "actively" supply the data. Indeed, in many cases website publishers and tracking companies (who partnered with the publisher) do not even inform website visitors in a meaningful way. (This lack of information is, by the way, a breach of data protection law and of the specific provisions in the e-Privacy Directive on cookies and similar files.[110]) Arguably, in situations in which consumers unknowingly supply data, their need for consumer protection would be even greater than in situations in which they submitted data knowingly. It is difficult to see why consumers in the two situations should be treated differently.[111] And as the European Data Protection Supervisor notes, the distinction between actively and not actively provided data conflicts with data protection law, which does not make such a distinction.[112]

Perhaps a more feasible route towards including services that are funded through targeted marketing within the scope of consumer law would have been to acknowledge that some services are not provided in exchange for a direct price or other kind of counter-performance, but are financed indirectly.[113] The draft Digital Content Directive should have clarified that consumer law should apply to such indirectly financed ("free") services. Ultimately, the goal must be to create an equal playing field with similar services that are offered in exchange for direct remuneration. It is difficult to see why consumers of commercial services that are financed in another way than direct remuneration should not receive protection under consumer law, despite the commercial value attached to immaterial forms of remuneration, such as data.

Another question is whether consumers of "free" services are entitled to the same level of protection as consumers who pay money. The draft Digital Content Directive is ambiguous on the question of whether consumers of "free" services can rely on similar protection to that consumers have when they pay money for services. At first sight the conformity test in Article 6 of the draft Directive seems to apply to free services.[114] This would mean that consumers of "free" digital content could complain about the lack of functionality, interoperability, accessibility, continuity, security, of the service. An open question is whether consumers could also complain, for instance, because the quality of personalized recommendations is below expectations.

However, the draft Digital Content Directive gives reason to doubt whether consumers of "free" services are entitled to a level of reasonable expectations which is similar to consumers who have paid a price (for the same service).[115] As Loos and Mak note, "the price the buyer is required to pay for the digital content will influence the reasonable expectations the buyer may have of the digital content".[116] On the other hand, personal data often function as an alternative rather than as a lower form of payment.[117] Following that line of reasoning, consumers may have reasonable expectations of digital content and services, even if they did not pay directly with money.

---

[109]The GDPR has strict rules on when consent is "freely given" and thus valid. See Art. 7 GDPR.
[110]See Art. 5(3) of the e-Privacy Directive (Directive 2002/58), as amended by Directive 2009/136. The Commission has published a proposal for an e-Privacy Regulation, which should replace the e-Privacy Directive: Proposal for a Regulation of the European Parliament and of the Council, concerning the respect for private life and the protection of personal data in electronic communications and repealing Directive 2002/58/EC (Regulation on Privacy and Electronic Communications), COM(2017) 10 final, <https://ec.europa.eu/digital-single-market/en/news/proposal-regulation-privacy-and-electronic-communications>.
[111]Mak, "The new proposal for harmonized rules on certain aspects concerning contracts for the supply of digital content", report for the European Parliament, 2016, at 9, <www.epgencms.europarl.europa.eu/cmsdata/upload/a6bdaf0a-d4cf-4c30-a7e8-31f33c72c0a8/pe__536.494_en.pdf>.
[112]EDPS, Opinion 4/2017, cited *supra* note 107, para 38.
[113]In a similar direction, EDPS Opinion 4/2017, ibid.
[114]This can be concluded, *a contrario*, from Art. 6(2)(a) of the proposed Digital Content Directive

A difficult question facing the draft Digital Content Directive is whether consumers of "free" services can rely on the same set of remedies as those for paid services in case of non-conformity, namely having the digital content brought into conformity free of charge, through a price reduction or the termination of the contract.[118] In this regard, the adequacy of the remedy (and the question of whether that remedy is disproportionate) would depend on the nature of the data and the service in question. For example, the appropriateness of the consumer rights concerning a social network would be different from the rights against the malfunctioning of a music streaming service.[119] The remedy of price reduction does not fit in the social network example (as it is difficult to express the value of data in monetary terms), and the conditions for being able to terminate are rather rigorous.[120] Similarly, the possibility to claim "economic damage" (Art. 14 draft Digital Content Directive) seems unsuitable for consumers of "free" services.[121]

What would be an adequate remedy if digital content or services are not in conformity? An example could be a smart phone app that collects more personal data than the app provider says in its contract with the consumer. The consumers' rights under data protection law would be of little help if the consumer is left with a non-functioning app. A more helpful remedy would probably be to reconfigure the app in a way that it ceases the contested data processing while retaining the same functionality. For the example of the smart phone app this could mean: if a judge finds that the collection of data is disproportionate, or exceeds what the parties have agreed to, the consumer can require that the app collects less data but must continue to provide the same functionality.

In conclusion, with the Unfair Commercial Practices Directive, and more recently the draft Digital Content Directive, consumer law has begun to consider the protection of consumers of so-called "free services". These are important first steps, but there is a need for further conceptualization and fine-tuning. New solutions, for instance regarding remedies, may be needed. In any case, we need a much better understanding of the implications of including "free" services under the scope of consumer law; section 5 of this article offers some initial thoughts on this.

---

[115] Art. 6(2)(a) of the proposed Digital Content Directive.

[116] In this sense also Loos and Mak, "Remedies for buyers in case of contracts or the supply of digital content", report for the European Parliament, Committee on Legal Affairs, 2012, at 180 <ec.europa.eu/justice/consumer-marketing/files/legal_report_final_30_august_2011.pdf>.
Art. 12 of the proposed Digital Content Directive

[117] In this sense Mak, "The new proposal for harmonized rules on certain aspects concerning contracts for the supply of digital content", Report for the European Parliament, Citizen's Rights and Constitutional Affairs at the request of the JURI Committee, Jan. 2016 <www.epgencms.europarl.europa.eu/cmsdata/upload/a6bdaf0a-d4cf-4c30-a7e8-31f33c72c0a8/pe__536.494_en.pdf>.

[118] Art. 12 of the proposed Digital Content Directive.

[119] Art. 12(1)(a) of the proposed Digital Content Directive.

[120] Art. 12(5) of the proposed Digital Content Directive.

[121] Critical on this restriction to economic damage also Metzger, op. cit. *supra* note 92, at 6; Spindler, "Verträge über digitale Inhalte – Haftung, Gewährleistung und Portabilität. Vorschlag der EU-Kommission zu einer Richtlinie über Verträge zur Bereitstellung digitaler Inhalte", (2016) *Multimedia und Recht*, 219-224, 222-223.

4.3. *Identifying unfair terms*

The Unfair Contract Terms Directive aims to protect consumers against unfair clauses included in pre-formulated contracts.[122] The Directive limits the possibilities for traders to impose contract terms upon consumers.[123] More generally, consumer law aims to promote fairness and to balance rights and obligations. Typically, those obligations relate directly to the subject matter of the contract (e.g. what quality or services users are entitled to expect, within which time frame, what obligations to maintain, etc.). But consumer law scholars Wilhelmsson and Willet explain that fairness rules in contract law can also be used to include other societal policies or entitlements from fundamental rights in the assessment of fairness.[124] The authors mention as an example of particularly problematic terms those that "impact the private sphere of life."[125]

The Unfair Contract Terms Directive declares that a contractual clause "shall be regarded as unfair if, contrary to the requirement of good faith, it causes a significant imbalance in the parties' rights and obligations arising under the contract, to the detriment of the consumer."[126] That provision could give courts room to consider, for instance, privacy and personal data related interests.

The fairness test in the Unfair Contract Terms Directive applies to "all contracts concluded between sellers or suppliers and consumers".[127] This control mechanism is not limited to specific types of contracts such as sales or services contracts. The Directive applies to contracts concluded by electronic means, for the provision of digital content and services and irrespective of the counter-performance.[128] The Directive potentially applies to a broad range of online transactions, such as downloading a "free" app from the iStore, buying a newspaper article at Blendle, and subscribing to an online advertising-funded music streaming service.

Under the Unfair Contract Terms Directive, the main subject matter of the contract and the adequacy of the price are excluded from the unfairness test, as long as these terms are in plain, intelligible language.[129] In many European countries, courts are not allowed to assess the fairness of the price when assessing the fairness of a contract. But if personal data are seen as "payment" for using, for instance, a social network site, the rules regarding "prices" could be applied. However, in the draft Digital Content Directive, data as counter-performance do not constitute the "price". If data are not considered as a price in the sense of the Unfair Contract Terms Directive, then the fairness control could apply to the conditions under which consumers are required to provide data provided to access a service.[130]

---

[122]Tenreiro, "The Community Directive on unfair terms and national legal systems: The principle of good faith and remedies for unfair terms", 3 *European Review of Private Law* (1995), 273–284.

[123]Unlimited contractual freedom would not fit in modern market conditions, in which consumers are often in a weaker position that traders. Weatherill, *EU Consumer Law and Policy*, 2nd ed. (Edward Elgar, 2014), p. 144.

[124]Wilhelmsson and Willet, "Unfair terms and standard form contracts" in Howells, Ramsay and Wilhelmsson (Eds.), *Handbook of Research on International Consumer Law* (Edward Elgar, 2010), pp. 158-191, 159-160. This is not the place to discuss the general positions on contract law on whether fairness considerations or considerations of party autonomy should serve as a point of departure. While in some countries there seems to be a focus on fairness considerations, in others, particularly in common law countries, party autonomy may trump, see Wilhelmsson and Willet, ibid. See also Collins, "Utility and rights in common law reasoning: Rebalancing private law through constitutionalization", LSE Law, Society and Economy Working Papers 6/2007, at 19. Collins points out that it might be necessary to translate e.g. the constitutional conception of privacy into a concept that fits better the realities of a relationship between private actors (rather than the State-citizen relationship).

[125]Wilhelmsson and Willet, ibid., p. 162.

[126]Art. 3(1) of the Unfair Contract Terms Directive.

[127]Recital 10 of the Unfair Contract Terms Directive.

[128]See Art. 3(1) of the Unfair Contract Terms Directive. About the conditions under which contract law applies to online contracts, Loos and Luzak, "Wanted: A bigger stick", 39 *Journal of Consumer Policy* (2016), 63–90.

[129]Art. 4(2) of the Unfair Contract Terms Directive stipulates that the "[a]ssessment of the unfair nature of the terms shall relate neither to the definition of the main subject matter of the contract nor to the adequacy of the price and remuneration, on the one hand, as against the services or goods supplies in exchange, on the other, in so far as these terms are in plain intelligible language".

In addition, several Member States do allow courts to examine the main obligations under the contract.[131] The European Court of Justice confirmed that the minimum harmonization level of the directives allows Member States to offer a higher level of protection to consumers.[132] Therefore, nothing prevents national laws allowing national courts to look at all the factual elements of a contract, including for instance price and counter-performance.[133]

Consumer law could thus be an important instrument to assess the fairness of terms and conditions regarding personal data, and the fairness of the conditions under which consumers agree to the processing of personal data in a commercial relationship. To return to our example of the torch app, consumer law's fairness test could be used to interpret data protection law's data minimization and purpose limitation principles. But data protection law could also provide an additional benchmark to assess the fairness of contractual conditions. For instance, a contract could be considered unfair if it breaches data protection law's data minimalization, but also e.g. security or privacy by default requirements.

Perhaps consumer law's fairness test could also be used to limit the abuse of consent as a legitimate ground for data processing. Consumer law could thereby provide a response to the increasing criticism that "consent" only protects consumers to a limited extent, as consumers often consent without reading the terms, or are left with little choice but consenting.[134] In the context of consumer law's fairness test it may not even matter if that information is personal data or not, thereby avoiding definitional questions about the scope of data protection law. Consumer law also provides for more flexibility as it also allows courts to consider the value that users get in return.

Consumer organizations have used consumer law on several occasions to scrutinize the fairness of the terms and conditions of companies that collect personal data, such as Google and Facebook.[135] The Federation of German Consumer Organizations has extensive experience in tackling unfair clauses of Facebook's and Apple's terms of use. Between 2009 and 2015 the Federation of German Consumer Organizations filed four injunctions against Facebook.[136] Two of these injunctions concern primarily the interplay between consumer and data protection laws. In one of these German Facebook cases, the Berlin Court of Appeal confirmed that data protection law provisions must be regarded as consumer protection provisions in the meaning of the German Act on Injunctive Relief (*Unterlassungsklagengesetz*).[137] This judgment has two consequences. First, consumer organizations can bring cases for data protection infringements. Second, the rules on unfair commercial practices or unfair contract terms can be applied to situations that concern personal data processing. In a case against Apple, the District Court of Berlin decided that the privacy policies of iTunes can be considered standard terms, and must comply with consumer law requirements regarding the clarity, specificity, and fairness of such standard clauses.[138] The court assessed eight clauses and found them unlawful, including a clause in which consumers had no option but to accept the sharing of personal data with third parties. According to the Berlin Court, those eight clauses constitute a significant imbalance between the parties' rights and obligations.[139]

---

[130]Arriving at the same conclusion, see Loos and Luzak, op. cit. *supra* note 128, at 67. The applicability of contract law is even less problematic if data are provided not in exchange for the service (because the service is remunerated), but as part of the contractual obligations of the consumer.

[131]Including Denmark, Greece, Spain, Luxembourg, Finland, Latvia, Malta, Portugal, Sweden and Slovenia, see Schulte-Nölke, Twigg-Flesner and Ebers (Eds.), *EC Consumer Law Compendium: The Consumer Acquis and its transposition in the Member States* (Sellier, 2008).

[132]Case C-484/08, *Caja de Ahorros y Monte de Piedad de Madrid* v. *Asociación de Usuarios de Servicios Bancarios (Ausbanc)*, EU:C:2010:309.

[133]Cámara Lapuente, *El Control de las cláusulas "abusivas" sobre elementos esenciales del contrato* (Thomson-Aranzadi, 2006), p. 98.

[134]Also under data protection law there is debate on whether consent can legitimize disproportionate data processing. E.g., Rouvroy and Poullet argue that: "even in case of unambiguous consent … it may be possible to declare the processing illegitimate if that processing is disproportionate", Rouvroy and Poullet, p. 73. See also Bygrave and Schartum, and Gellert and Gutwirth, at 527; all op. cit. *supra* note 35.

[135]See also Rott, "Data protection law as consumer law: How consumer organisations can contribute to the enforcement of data protection law", 6 *Journal of European Consumer and Market Law* (2017), 113.

[136]The cases concern different features of Facebook services, including the transfer of data to third parties via the app-centre, the "friends finder" tool, the advertising of the services as "free" and the transfer of data from WhatsApp to Facebook in the post-merger.

In 2016, the Norwegian Consumer Council scrutinized contract terms of popular apps such as Facebook, Instagram, LinkedIn, and Twitter under legislation on unfair terms and data protection.[140] The Norwegian Consumer Council noted, among other breaches of the law, that some apps use vague language when describing their personal data use, and seek excessive permissions to access personal data stored in consumer phones. The Norwegian Consumer Council complained to the Norwegian Consumer Ombudsman that the terms and conditions of the dating app Tinder violated the unfair contract terms legislation.[141] The Consumer Council complained, among other things, that the terms and conditions became binding by tacit agreement i.e. by using the service. In this sense, Tinder did not give consumers the possibility to acquaint themselves with the terms and conditions and the subsequent changes. The new contract allowed Tinder to access the consumer's information available in other apps, such as Facebook and Instagram. In a second complaint, the Norwegian Consumer Council denounced the app Runkeeper for continuing to track users even when the app was not activated.[142] These cases demonstrate the role that consumer law could play in the scrutiny of terms related to personal data collection and processing. The application of consumer law could also be extended to privacy notices included in the terms of use of connected products and the Internet of Things. In a recent co-coordinated action, US and EU consumer groups asked consumer agencies and data protection authorities to look at data protection and consumer law infringements of connected toys.[143]

---

[137] Gesetz über Unterlassungsklagen bei Verbraucherrechts- und anderen Verstößen (Unterlassungsklagengesetz - UKlaG), 26 Nov. 2001, as amended by Art. 3 G, 28 Apr. 2017 I 969 (Nr. 23).
[138] Landgericht Berlin, Judgment of 30 Apr. 2013, 15 O 92/12 <www.vzbv.de/sites/default/files/downloads/Apple_LG_Berlin_15_O_92_12.pdf>.
[139] Ibid, at 10. Similarly, the French consumer associations UFC-Que Choisir took Google, Facebook and Twitter to a civil court for infringing several French and European laws, including consumer law, data protection law, and copyright law. The three cases are still pending in the first instance court of Paris. UFC-Que Choisir, "L'UFC-Que Choisir attaque les réseaux sociaux et appelle les consommateurs à «garder la main sur leurs données»", 25 March 2014, <www.quechoisir.org/action-ufc-que-choisir-donnees-personnelles-l-ufc-que-choisir-attaque-les-reseaux-sociaux-et-appelle-les-consommateurs-a-garder-la-main-sur-leurs-donnees-n11951/>.
[140] Forbrukerrådet, "APPFAIL – Threats to Consumers in Mobile Apps", March 2016 <fbrno.climg.no/wp-content/uploads/2016/03/Appfail-Report-2016.pdf>.
[141] Forbrukerombudet, "Complaint regarding unfair contractual terms in the Terms of Use for the mobile application Tinder", March 2016 <fbrno.climg.no/wp-content/uploads/2016/03/20160302-Complaint-Tinder.pdf>.
[142] Forbrukerombudet, "Runkeeper tracks users when the app is not in use", May 2016 <www.forbrukerradet.no/side/runkeeper-tracks-users-when-the-app-is-not-in-use/>.
In this case the Norwegian Consumer Council complained to the Norwegian Data Protection Authority, rather than to the Consumer Ombudsman (unlike the Tinder complaint). The difference is due to the legal basis used to attack the specific infringement, either consumer law or data protection law.
[143] BEUC, "Consumer organisations across the EU take action against flawed internet-connected toys", 6 Dec. 2016, <www.beuc.eu/publications/consumer-organisations-across-eu-take-action-against-flawed-internet-connected-toys/html>; BBC News, "Call for privacy probes over Cayla doll and i-Que toys", 6 Dec. 2016,

In conclusion, consumer law can be a useful tool to safeguard the overall balance in the commercial relationship between consumers and suppliers. Consumer law can also be used to assess the fairness of situations in which companies require consumers to consent to the processing of disproportionate amounts of data, to sharing of data with third parties, etc.

4.4. *Identifying unfair commercial practices*

The Unfair Commercial Practices Directive can also help to assess whether companies deal fairly with personal data.[144] The Unfair Commercial Practices Directive prohibits unfair commercial practices.[145] The Directive defines a commercial practice as "any act, omission, course of conduct or representation, commercial communication including advertising and marketing, by a trader, directly connected with the promotion, sale or supply of a product to consumers." Does the Unfair Commercial Practices Directive apply if a consumer consents to his or her personal data being processed? If consenting to personal data processing is seen as a transactional decision,[146] then the Unfair Commercial Practices Directive could help to assess the fairness of the conditions under which users are required to agree to the collection and use of their personal data - e.g. take-it-or-leave-it choices, misinforming about the functionality of the service if consumers do not agree, etc. The Directive would allow us to assess such practices in the light of particularly vulnerable consumers, such as children and elderly people, but also, perhaps, the profiled user (see further below).

According to the Commission, both the wording of the provision and the case law of the ECJ suggest a broad interpretation of the notion of transactional decisions, in the sense of "any decision directly related to that decision".[147] The Berlin Court of Appeals stated that a consumer's decision to agree to data processing as a pre-condition for being able to use a service could be considered a transactional decision in the sense of the Unfair Commercial Practices Directive.[148] This suggests that the Directive can be used to scrutinize the fairness of the conditions under which the user is asked to agree to data processing, whether the user has been properly informed, was not put under undue pressure, has been misled etc. According to the Berlin court, the decision to provide personal data for personalized advertising is a transactional decision of the consumer. The court argues: "The decision whether or not the consumer agrees to being subjected to advertising persuasion – and personally targeted persuasion in particular – must be part of the free autonomous choice of the consumer".[149] Accordingly, says the court, the decision to consent to personal data processing is not only relevant for data protection law, but is also a matter of consumer protection, and hence a transactional decision in the sense of the Unfair Commercial Practices Directive.[150]

---

<www.bbc.com/news/technology-38222472>; Forbrukerombudet, "Connected toys violate European consumer law", 6 Dec. 2016, <www.forbrukerradet.no/siste-nytt/connected-toys-violate-consumer-laws>.
[144]Kannekens and Van Eijk, "Oneerlijke handelspraktijken: alternatief voor privacy handhaving", 4 *Mediaforum* (2016), at 24.
[145]Art. 5(1) of the Unfair Commercial Practices Directive.
[146]Art. 2(k) of the Unfair Commercial Practices Directive describes a "transactional decision" as any decision take by a consumer concerning whether, how and on what terms to purchase, make payment in whole or in part for, retain or dispose of a product or to exercise a contractual right in relation to the product, whether the consumer decides to act or do refrain from acting.
[147]Commission guidance cited *supra* note 72, at 37, referring to Case C-281/12, *Trento Sviluppo*, paras. 36 and 38.
[148]Kammergericht Berlin, Judgment of 24 Jan. 2014, 5 U 42/12; available at <www.vzbv.de/sites/default/files/downloads/Facebook_II__Instanz_AU14227-2.pdf>.
[149] "Denn es muss in der freien Entscheidung des Verbrauchers liegen, inwieweit er sich einer Verführung durch Werbung – insbesondere einer individuelle zielgerichteten – assetzen will" (translation by the authors).
[150]Kammergerciht Berlin, Judgment of 24 Jan. 2014, 5 U 42/12, at 33.

Under the Unfair Commercial Practices Directive, sellers must act in accordance with professional diligence. If the Directive applies, compliance with data protection law can be part of the professional diligence that sellers owe consumers. The decision of the Berlin Court of Appeals,[151] later confirmed by the Federal Supreme Court,[152] found that the "Find Friends" feature of Facebook violates the requirement to obtain the user's consent under German data protection law because Facebook members do not consent to the data collection after clicking the "Find Friends" button[153] The court considered this situation an infringement of the German law implementing the Unfair Commercial Practices Directive.[154] The characterization as an unfair commercial practice is due to the infringement of a "statutory provision that is also intended to regulate market behaviour in the interest of market participants."[155] Consequently, a data protection infringement could simultaneously amount to a consumer law infringement.

The considerations of the Berlin Court of Appeals touch upon another aspect, namely the potential persuasiveness of behavioural targeting (or what the court calls the "hohe Effizienz einer individuell auf den jeweiligen Verbraucher zugeschnittenen Werbung"[156]) and how this relates to the provisions under the Unfair Commercial Practices Directive. The Unfair Commercial Practices Directive may help to provide guidance on the fairness of behavioural targeting, because one of the Directive's objectives is safeguarding the consumers' autonomous decision-making process. The Directive's provisions against deception, unfair restrictions of consumer choices and aggressive marketing practices serve this purpose. As Howells explains, a central element of the provisions about aggressive practices is protecting the consumer's freedom to choose; he points to the thin line between advertising as a form of (legitimate) persuasion and the exercise of undue influence or even coercion.[157] Hence, the Unfair Commercial Practices Directive could be relevant in situations in which profiling is used to influence consumers' decision-making.[158]

More generally, there is a privacy protection element to some provisions of unfair commercial practice law. The rules about unfair commercial practices, coercion and harassment have typically been applied to situations of doorstep selling or phoning people at their homes, i.e. situations in which the personal autonomy and privacy of consumers is at stake.[159] Nowadays, consumers are not only approached by merchants on their doorsteps, but maybe even more directly in their private realm, namely on phones, smart TVs and other devices in their homes, and sometimes even on their wrist or in their breast pocket. With the Internet of Things and the proliferation of smart devices that are also being used to communicate marketing messages, the Unfair Commercial Practices Directive might play a role in protecting consumers against privacy-intrusive or unfair persuasion.[160]

---

[151] Kammergericht Berlin, Judgment of 24 Jan. 2014, 5U 42/12.
[152] Bundesgerischtshof, Judgment of 14 Jan. 2016, I ZR 65/14
[153] This is a tool allowing Facebook users to find other users by aggregating data from mailing services such as Yahoo!, Gmail or Skype.
[154] Gesetz gegen den Unlauteren Wettbewerb – UWG.
[155] Art. 3(a) UWG.
[156] Kammergerciht Berlin, Judgment of 24 Jan. 2014, 5 U 42/12, at 34.
[157] Howells, "Aggressive commercial practices" in Howells, Micklitz and Wilhelmsson (Eds.), *European Fair Trading Law* (Ashgate, 2006), p. 168.
[158] Howells, ibid., pp. 167-195, 178; Howells, Micklitz and Wilhemsson, "Towards a better understanding of unfair commercial practices", 52 *International Journal of Law and Management* (2009), 69-90.
[159] In the US, the Federal Trade Commission has played an important role in protecting privacy. As the US scholars Solove and Hartzog note, "FTC privacy jurisprudence has become the broadest and most influential regulating force on information privacy in the United States – more so than nearly any privacy statute or any common law tort". Solove and Hartzog, "The FTC and the new common law of privacy", (2014) *Columbia Law Review*, 584-676.

4.4. *Profiling and consumer vulnerability*

There is some fear that online profiling could be used to manipulate people. Personalized ads could be used to exploit people's weaknesses or to charge people higher prices. Calo worries that in the future, companies could find people's weaknesses by analysing massive amounts of information about their behaviour: "digital market manipulation".[161] With modern personalized marketing techniques, "firms can not only take advantage of a general understanding of cognitive limitations, but can uncover and even trigger consumer frailty at an individual level."[162] For example, a company could target ads to somebody when they are tired, or when they are easy to persuade for another reason. Companies could tailor messages for maximum effect. Such worries may not be completely unfounded. One marketing company suggests advertising beauty products on Mondays; its press release says: "New beauty study reveals days, times and occasions when U.S. women feel least attractive".[163] And, reportedly, "Facebook showed advertisers how it has the capacity to identify when teenagers feel 'insecure', 'worthless' and 'need a confidence boost'".[164] A company could also learn what kinds of arguments convince an individual to buy a product. Does somebody react to discounts, or to phrases such as "special offer, only today"?[165]

To protect people effectively, the law must offer the flexibility to consider individual characteristics and vulnerabilities, including those that are the result of digital profiling and personalized marketing. People are heterogeneous, and have different characteristics, needs, preferences, and, perhaps most importantly, vulnerabilities.[166] Consumer law has a long tradition of conceptualizing the consumer. After all, the level of protection needed depends on who the consumer is. Is it the archetypical vulnerable consumer, finding themselves in a situation of structural weakness in relationship to the supplier, and therefore requiring a more protective regime? Or is it the empowered consumer, who typically knows where their preferences and weaknesses lie, and who acts as an active market participant? The ECJ has made it clear that the reference consumer is the empowered consumer, or, in the terms of the ECJ, the "reasonably informed, observant and circumspect" consumer.[167] The Court's concept of the consumer is reflected throughout European consumer law.[168]

---

[160]For a first analysis see Helberger, "Profiling and targeting in the Internet of Things: A new challenge for consumer protection" in Schulze and Staudenmayer (Eds.), *Digital revolution: Challenges for Contract Law in Practice* (Baden Baden, 2016), pp.135-161.
[161]Calo, "Digital market manipulation", 27 *George Washington Law Review* (2013), <ssrn.com/abstract=2309703>.
[162]Ibid., at 1.
[163]PRnewswire, "New beauty study reveals days, times and occasions when U.S. women feel least attractive" (2013), <www.prnewswire.com/news-releases/new-beauty-study-reveals-days-times-and-occasions-when-us-women-feel-least-attractive-226131921.html>; see also Rosen, op. cit. *supra* note 12.
[164]The Guardian, "Facebook told advertisers it can identify teens feeling 'insecure' and 'worthless'", 1 May 2017, <www.theguardian.com/technology/2017/may/01/facebook-advertising-data-insecure-teens>.
[165]Kaptein, Eckles, and Davis, "Envisioning persuasion profiles: Challenges for public policy and ethical practice", 18 *Interactions* (2011), 66-69, 66; see also Kaptein, Lacroix and Saini, "Individual differences in persuadability in the health promotion domain" in Plough, Hasle and Oinas-Kukkonen (Eds.), *Proceedings of 5th International Conference on Persuasive Technology: PERSUASIVE 2010* (Springer, 2010), pp. 82-93.
[166]The European Consumer Consultative Group in its opinion on consumers and vulnerability suggested the need to introduce a new approach to the way to conceive consumer vulnerbly in a critic to the current notion of average consumer: "Consumer vulnerability can be determined in the first place by personal characteristics such as age, disability, or income ('the personal dimension'). Because these socio-economic factors have an impact on the way consumers act in general, the 'personal dimension of consumer vulnerability' also triggers the question on whether the present 'horizontal approach', based on the notion of 'average consumers', is fit to protect all consumers", European Consumer Consultative Group, "Opinion on Consumers and Vulnerability" (2013) <ec.europa.eu/consumers/archive/empowerment/docs/eccg_opinion_consumers_vulnerability_022013_en.pdf>.

Nevertheless, consumer law contains some flexibilities to accommodate vulnerabilities of individual consumers or consumer groups. The Unfair Commercial Practices Directive expresses this flexibility in the concept of the vulnerable consumer. For example, according to the Directive, commercial practices that are likely to materially distort the economic behaviour only of a clearly identifiable group of consumers shall be assessed from the perspective of the average member of that group.[169] This is typically the case for (groups of) consumers who are particularly vulnerable to the practice or the underlying product because of their mental or physical infirmity, age or credulity. If one defines "vulnerability"[170] as the "limited ability to deal with commercial practices",[171] at what point does profiling-based marketing turn the normal, "average" consumer into a vulnerable one? With personalized marketing, companies could automatically adapt advertisements to (inferred) characteristics, biases and weaknesses of individual consumers. Possibly, in a digital environment, new groups of vulnerable consumers need to be identified. This could be, for instance, particularly active online users who leave a correspondingly large data-footprint, "quantified self" consumers who use smart devices to track their own behaviour, or those that are particularly perceptible to digital market manipulation. Because of the possibility to target individuals, there needs to be more legal attention for individual characteristics and vulnerabilities. Further normative and empirical research is needed in this area.[172] In sum, profiling may lead to new forms of consumer vulnerability in the sense of the Unfair Commercial Practices Directive; that Directive could help to mitigate possible harms.

## 5. A perfect match?

In a modern economy, the collection and processing of personal data affects people not only as data subjects and holders of the fundamental right to privacy, but also as consumers. We have shown that the application of consumer protection law to data-related commercial practices can add to the protection offered by data protection law. Together, data protection and consumer law will lay down some ground rules for the modern economy. As consumers, people must, on a daily basis, assess the fairness and desirability of deals that involve the processing of their personal data.¹⁷³ For instance, for so-called Internet of Things products, companies often integrate products, services, and personal data collection in one contract.¹⁷⁴

---

[167]Case C-210/96, *Gut Springenheide GmbH, Rudolf Tusky and Oberkreisdirektor des Kreises Steinfurt Amt für Lebensmittelüberwachung*, EU:C:1998:369.
[168]See Rinkes, "Europees consumentenrecht" in Hondius (Ed.), *Handboek Consumentenrecht. Een overzicht van rechtspositie van de consument*, (Uitgeverij Paris, 2006), p. 36. See also Schebesta and Purnhagen, "The behaviour of the average consumer: A little less normativity and a little more reality in CJEU's case law? Reflections on Teekanne" (6 June 2016), <ssrn.com/abstract=2790994>.
[169]Art. 3(3) of the Unfair Commercial Practices Directive.
[170]Commission, Consumer vulnerability across key markets in the European Union: Final report (2016), <ec.europa.eu/consumers/consumer_evidence/market_studies/vulnerability/index_en.htm>, at 39.
[171]Duivenvoorde, "The protection of vulnerable consumers under the Unfair Commercial Practices Directive", 2 *Journal of European Consumer and Market Law* (2013), 69-79, at 73.
[172]See Madden and Rainie, "Americans' attitudes about privacy, security and surveillance" (2015), <www.pewinternet.org/2015/05/20/americans-attitudes-about-privacy-security-and-surveillance/>. A very small number of consumers indicated having actively changed their behaviour to avoid being tracked, but they also found that many consumers are engaged in some form or other in more common or less technical privacy-enhancing measures. Comparable research for Europe is still scarce.

In this article, it was shown that data protection law and consumer law can usefully complement each other. Whereas data protection law looks primarily at the fairness of the collection and processing of personal data, consumer law broadens that perspective, and provides tools to assess the balance and fairness in the broader commercial relationship between consumers and data controllers. Consumer law can have an important role in balancing the (often unequal) negotiating power between consumers and providers of digital services, helping regulators and courts to develop and modernize the existing catalogue of fair and unfair practices. The rules on unfair commercial practices could help to protect consumers if companies used profiling to unfairly influence consumer decisions. Consumer law, and the flexibility of the fairness test in particular, could also add an extra safeguard against the use of consent as a means to legitimize data collection and processing, which puts consumers into a situation of imbalance *vis-à-vis* the supplier.

*Vice versa*, data protection law can inform the interpretation and the development of consumer law, and thereby help to adjust consumer law practice better to the demands of the modern economy, where personal data processing plays a large role. Wendehorst argues that consumer law should require companies to implement privacy by design - a data protection principle - in Internet of Things products.¹⁷⁵ If consumer law applied, people would not only be able to exercise their rights under data protection law, but they would also be able to, for instance, return a product or demand their money back. Similarly, data protection law can help to identify certain unfair commercial practices. For example, suppose that a company abuses behavioural targeting to exploit the vulnerability of certain consumers. People with health-problems could be targeted with ads for certain products. Data protection law could help to identify such problematic advertising practices, as data protection law has stricter rules for "special categories" of data, such as data related to people's health.¹⁷⁶

However, consumer law and data protection law are not a perfect match – at least not yet. In the following section, we point to some of the conceptual growing pains of a more integrated vision on data protection and consumer law, such as revealed in the draft Digital Content Directive. We also ponder on the broader implications and risks of such a vision.

5.1. *Consistency*

Consumer law and data protection law are two different fields, with different legal traditions, concepts and objectives.¹⁷⁷ It remains to be seen to what extent notions such as harm, fairness, damages or data will be applied consistently across the two fields of law. The draft Digital Content Directive is only the start of a discussion on the best ways of conceptualizing and operationalizing the interaction between the two different areas of law.

Several aspects of the draft Digital Content Directive are unclear. For example, it is unclear what quality and functionality consumers may expect of services that are rendered in return for data. The ECJ emphasizes that people are entitled a "high" level of protection of their personal data.¹⁷⁸ And yet, under the draft Digital Content Directive, consumers who acquire

---

¹⁷³See United Nations Guidelines on Consumer Protection, <unctad.org/en/Pages/DITC/CompetitionLaw/UN-Guidelines-on-Consumer-Protection.aspx>.
¹⁷⁴Wendehorst, op. cit. *supra* note 92.
¹⁷⁵Wendehorst, op.cit. *supra* note 92, at 15.
¹⁷⁶ See Art. 9 of the GDPR.
¹⁷⁷See sections 2 and 3 *supra*.
¹⁷⁸Case C-131/12, *Google Spain* v. *Agencia Española de Protección de Datos (AEPD) and Mario Costeja González*, EU:C:2014:317, para 66. See also Recital 6 and 10 GDPR. See also Junker's State of the Union Address

services in exchange for data seem not to be entitled to expect a comparable standard of functionality and quality, as compared to situations where products and services have been paid for with money.[179]

Another source of confusion is the rights granted to consumers. The draft Digital Content Directive suggests a requirement for sellers to refrain from using data given as a counter-performance once a contract has been terminated.[180] That right to require non-use of data is probably meant as an equivalent of the right to ask money back. But it is unclear what the provision adds to rights that people have under data protection law. For instance, data protection law allows people to withdraw their consent, and grants people a right to erase personal data, and to object to processing.[181] More generally, the effectiveness of the right to require non-use is open to question.[182]

Yet another source of inconsistency is that the draft Digital Content Directive refers to "data" and not only to "personal data". The use of the word "data" could make the scope of the draft Digital Content Directive broader (in certain respects) than data protection law. But it is unclear what types of non-personal data consumers would provide to online service providers or whether such data would cover data generated by the consumer when using the devices or digital content.[183]

### 5.2. *Conflicts and contradictions*

At times, the provisions in consumer law and in data protection law seem to be in conflict. For example, whereas the draft Digital Content Directive explicitly acknowledges the provision of personal data as a possible counter-performance for rendering services,[184] under the GDPR, consent must be "freely given" to be valid. If a company requires consumers to provide data beyond what is strictly necessary for the performance of a contract, such consent may not be "freely given" under the GDPR.[185] Article 7(4) of the GDPR states: "When assessing whether consent is freely given, utmost account shall be taken of whether, *inter alia*, the performance of a contract, including the provision of a service, is conditional on consent to the processing of personal data that is not necessary for the performance of that contract."[186] If a company processes personal data on the basis of invalid consent, the company may not have a legal basis for the processing. If the processing is unlawful, though, how can it be the basis for a contractual interaction?

Perhaps the draft Digital Content Directive should be interpreted as follows. Data can be considered as a counter-performance for the purpose of the Digital Content Directive, but the mandatory rules of the GDPR always apply. In other words, the inclusion of data as a counter-performance (in consumer law) does not legitimize processing of personal data in breach of the GDPR principles. Rather, data protection law could add to the interpretation of

---

2016: Towards a better Europe - a Europe that protects, empowers and defends, Strasbourg, 14 Sept. 2016. Online available at: <europa.eu/rapid/press-release_SPEECH-16-3043_en.htm>.
[179]See section 4.2. *supra*.
[180]Art. 13(2)(b) draft Digital Content Directive.
[181]See Art. 7(3), Art. 17, Art. 21 of the GDPR.
[182]See (critical) also Mak, op. cit. *supra* note 117, at 9.
[183]Art. 7(3) of the GDPR states: "When assessing whether consent is freely given, utmost account shall be taken of whether, *inter alia*, the performance of a contract, including the provision of a service, is conditional on consent to the processing of personal data that is not necessary for the performance of that contract." See also EDPS, Opinion 4/2017, cited *supra* note 107.
[184]Art. 3(1) Draft Digital Content Directive.
[185]See also EDPS, Opinion 4/2017, para 55, cited *supra* note 107.
[186] See on Art. 7(4) GDPR, Kostić and Vargas Penagos, "The freely given consent and the 'bundling' provision under the GDPR", (2017) *Computerrecht,* 2017-4, pp. 217-222.

consumer law. Metzger, for example, suggests interpreting the GDPR "as an appeal to contracting parties and courts to pay special intention to the voluntary nature of the consumer's consent when consent is given within the framework of a contractual relationship."[187] On the contrary, Langhanke and Schmidt-Kessel suggest (perhaps controversially) interpreting Article 7(4) GDPR in a more restrictive manner in the sense that it should not apply to contracts where "a whole obligation under the contract builds on commercialization of personal data".[188] These two views represent two ends of the spectrum of possible ways of weighing the valuations in consumer law vs. data protection law. As argued below, Metzger's restrictive interpretation of consumer law in the light of data protection law seems to be the right approach, given the individual and societal importance of the right to personal data protection.

To give a further example: it is still far from clear to what extent consumers can meaningfully commit to providing data as counter-performance if, according to data protection law, they have the right to withdraw their consent to data processing at any moment in time. The implications for the contractual relationship between consumers and service providers are not yet well understood.[189]

Under consumer and contract law, consumers do not only have rights, they also have obligations.[190] For example, to be able to exercise their rights, consumers must check products and services, notify providers of potential failures and instances of non-compliance, and fulfil their part of a contractual obligation. It is unclear how far contractual obligations could go. Could consumers be required by contract law to provide personal data?[191] Could consumers have a contractual obligation to keep data up-to-date for the duration of the contract? What happens if a consumer withdraws his or her consent, a right the consumer always has under the GDPR? Could the provision of false data (e.g. false name, age or address) constitute a breach of contract?

5.3. *Should the law regard personal data as a commodity?*

An important caveat is in order. Acknowledging that data can be a counter-performance under consumer law does not mean that personal data are equivalent to money. Data are indeed valuable assets for companies.[192] But fundamental rights, such as the right to privacy and to personal data protection, which also have a societal dimension, should not be downgraded to mere individual consumer interests.[193]

In the European Union, a strong push for a "European Data Economy" is well under way.[194] The Commission has explained its vision of a "well-functioning and dynamic data

---

[187]Metzger, op. cit. *supra* note 92, at 5. Metzger focuses on Art. 7(4) of the GDPR in that sentence. On the contrary, Langhanke and Schmidt-Kessel suggest (perhaps controversially) interpreting Art. 7(4) GDPR in a more restrictive manner in the sense that it should not apply to contracts where "a whole obligation under the contract builds on commercialization of personal data". Langhanke and Schmidt-Kessel, op. cit. *supra* note 92, at 220.
[188] Ibid., at 220.
[189]In this sense also Metzger, ibid., at 6 (also pointing out, however, that such synallagmatic contracts are not unprecedented, at least not in German consumer law); Langhanke and Schmidt-Kessel, ibid., at 221.
[190]Heidbrink, Schmidt and Ahaus, *Die Verantwortung des Konsumenten: Über das Verhältnis von Markt, Moral und Konsum* (Campus, 2011), pp. 79 et seq.
[191]In this sense also Metzger, op. cit. *supra* note 92, at 6.
[192]Several mergers and acquisitions were meant, at least in part, to acquire data or data collection possibilities. See e.g. Google / DoubleClick, Facebook / WhatsApp and Microsoft / LinkedIn.
[193]See Arts. 7 and 8 EU Charter of Fundamental Rights.
[194]Defined by the European Commission as follows: "The data economy measures the overall impacts of the data market – i.e. the marketplace where digital data is exchanged as products or services derived from raw data – on the economy as a whole. It involves the generation, collection, storage, processing, distribution, analysis, elaboration, delivery, and exploitation of data enabled by digital technologies (European Data Market study, SMART 2013/0063, IDC, 2016)", Commission, Communication to the European Parliament, the Council, the

economy" that "requires the flow of data in the internal market to be enabled and protected."[195] In the Commission's view, data are economic assets, and monetizing data is a precondition for the prospering of the EU. In such a political climate, it is even more important to emphasize that neither laws nor policies should fuel the idea that people can renounce their rights in exchange of services.

The draft Digital Content Directive tries to accommodate these concerns, at least in part. The draft Directive qualifies data collected or provided in exchange of a service as a "counter-performance", and not as a price. However, this approach might not be sufficient to cover services that can be financed indirectly (for instance through behavioural advertising). In a similar vein, the European Data Protection Supervisor, while supporting the extension of consumer law to cover these digital services, raises concerns about the consideration of personal data as a commodity.[196] In line with the suggestion we made in this article, the European Data Protection Supervisor suggests using the definition of services of the E-commerce Directive,[197] which would cover different modes of remuneration, instead of using the term "counter-performance".

5.4. *Open questions*

Many questions are still open regarding the interplay between consumer law and data protection law. What are adequate remedies when consumers do not pay money for services and traders capture consumer data? How can one assess the value of personal data in contractual exchange relationships, and what are the limits that fundamental rights doctrine poses? Other aspects that merit further exploration are the institutional and organizational settings, and questions such as the levels of expertise in tech-savviness in data protection authorities *vis-à-vis* consumer protection authorities, the impact of differences in jurisdiction and competences. Also, it may be fruitful to engage in comparative research, comparing the situation in the US (where the Federal Trade Commission developed principles for the fair use of personal information on the basis of consumer law) and Europe (where most data-related issues fall primarily under the scope of data protection law). These are all pieces in a large puzzle that can ultimately lead to the development of a more comprehensive vision on "data consumer law", and the protection of fairness and fundamental rights in digital consumer markets.

**6.     Conclusion**

Consumer law and data protection law can usefully complement each other and offer people more complete protection in modern markets. By focusing on the economic transaction between consumers and traders, consumer law goes beyond data protection law in some aspects. Consumer law focuses on the impact on individual (economic) behaviour and the choices of consumers. It introduces more flexibility to assess the fairness of services that monetize personal data. It could help to protect consumers against unfair profiling and persuasion practices. And consumer law provides contractual remedies in case of a breach of contract or non-performance of the service by the service providers, for example because a service fails to respect the standard that is set by data protection law.

---

European Economic and Social Committee and the Committee of the Regions, Building a European Data Economy, COM(2017)9 final, at 2.
[195]Ibid, at 5.
[196]EDPS, Opinion 4/2017, cited *supra* note 107, at 7.
[197]E-commerce Directive 2000/31/EC, O.J. 2000, L 178/1.

The draft Digital Content Directive tries to update traditional consumer law by introducing remedies applicable to contracts involving exchange of personal data. The draft Directive introduces the new legal recognition of providing data as a consumer's "counter-performance" in a contractual relationship. This legal recognition could mean the start of using consumer law to assess the fairness of contracts regarding services that are presented to the consumer as "free". However, applying consumer law to deals regarding personal data should never be construed as a justification for using personal data as a commodity. Seeing personal data merely as tradeable goods would conflict with human rights.

Additionally, data protection law can inform the interpretation of consumer law. Using consumer rights, consumers should be able to challenge the excessive collection of personal data in take-or-leave-it situations. Disproportionate data processing is already illegal under protection law. Under consumer law, contractual clauses requiring consumers to disclose excessive amounts of personal data could also be unfair, because of the imbalance between the parties' rights and obligations which is to the detriment of the consumer and against good faith.

This article has argued that the interplay of data protection law and consumer protection law provides new ways to address situations of consumers detriment and unfairness in digital markets. This is but the beginning of an exploration of how the two areas of law interact. As was also demonstrated, the relationship between consumer law and data protection law still reveals inconsistencies and even contradictions. Nevertheless, and as was seen with the experience of consumer organizations who have successfully used consumer law to tackle data protection infringements, the interplay of data protection law and consumer protection law provides exciting opportunities.

\* \* \*